\documentclass[12pt,reqno]{amsart}
\usepackage{graphicx}
\usepackage{amssymb,amscd,amsbsy}
\usepackage{amsthm}

\newcommand{\noi}{\noindent}

\newcommand{\RF}{{\mathbb R}}
\newcommand{\CP}{{\mathbb C}{\mathbb  P}}
\def\ec{\epsilon^{\circ}}
\newcommand{\HH}{{\mathcal H}}%
\newcommand{\MM}{{\mathcal M}}
\def\PP{\mathcal   P}

\def\[{\left[}%
\def\]{\right]}
\def\({\left(}
\def\){\right)}
\def\l{\lambda}
\def\g{\gamma}
\newcommand{\pa}{\partial}
\newcommand{\eeq}{\end{equation}}
\newcommand{\beq}{\begin{equation}}
\newcommand{\bay}{\begin{eqnarray}}
\newcommand{\ey}{\end{eqnarray}}
\newcommand{\bey}{\begin{eqnarray*}}
\newcommand{\eey}{\end{eqnarray*}}

\newcommand{\Rat}{\operatorname{Rat}}
\newcommand{\R}{\operatorname{res}}

\newcommand{\tr}{\operatorname{trace}}

\newtheorem{thm}{\hspace{\parindent}Theorem}[section]

\newtheorem{lem}[thm]{\hspace{\parindent}Lemma}

\usepackage{amsfonts,amssymb}
\usepackage{amsthm}
\usepackage{amscd}
\usepackage{array}
\usepackage{color}
\usepackage{pstricks,pst-node}
\usepackage{pst-plot}
\usepackage{pst-solides3d}

\theoremstyle{remark}

\newtheorem*{rem*}{Remark}

\begin{document}

\newcommand{\vse}{\vspace{.2in}}
\numberwithin{equation}{section}

\title{\bf The  hierarchy  of Poisson brackets   for the open Toda lattice  and its' spectral curves}
\author[1]{K.L.  Vaninsky}
\begin{abstract}
We establish a  new  representation of the infinite  hierarchy of  Poisson brackets (PB)   for  the open Toda lattice in terms of its spectral curve.
For the classical Poisson bracket (PB) we give a representation  in the form of a contour integral of
some special Abelian differential (meromorphic one-form) on the spectral curve.
All higher brackets of the infinite hierarchy  are obtained by multiplication of the  one-form  by a power of the  spectral parameter.
\end{abstract}
\maketitle
\tableofcontents

\setcounter{section}{0}
\setcounter{equation}{0}
\section{Introduction.}

All  known  integrable hierarchies of equations like Toda lattice,  Camassa--Holm equation,  Korteveg de Vriez equation,  Nonlinear Schr\"{o}dinger equation, sine-Gordon equation, Landau-Lifshitz  equations are  Hamiltonian systems. In fact on their phase space $\MM$ there exists  a  finite or  infinite set of
commutative vector fields
$$
X_1, X_2,\hdots
$$
that are compatibility conditions for the Lax's equations. This vector fields can be written  with  the classical Poisson bracket $\{\;, \; \}_{\pi_0}$
and different      Hamiltonians   $\HH_1, \HH_2, \HH_3,...;$  as
$$
X_k=\{\;\;,\HH_k\}_{\pi_0},\qquad k=1,2,....
$$
In all these examples there exists a second bracket $\{\;, \; \}_{\pi_1}$   compatible with the first   $\{\;, \; \}_{\pi_0}$.
For the KdV  the bracket   $\{\;, \; \}_{\pi_1}$    is called the  Lenard-Magri bracket.    All vector fields can be written with respect to this   second bracket.

In this paper we study a Hamiltonian theory of the finite open Toda lattice.
The open finite  Toda lattice is a mechanical system of $N$--particles connected by
elastic  strings.  The Hamiltonian of the system is
$$
H=\sum\limits_{k=0}^{N-1}\frac{p_k^2}{ 2} +\sum\limits_{k=0}^{N-2}e^{q_k-q_{k+1}}.
$$
Introducing the classical Poisson bracket
\beq\label{cpb}
\{f,g\}_{\pi_0}=\sum\limits_{k=0}^{N-1} \frac{\pa f}{\pa q_k} \frac{\pa g}{ \pa p_k}-
                                \frac{\pa f}{ \pa p_k} \frac{\pa g}{\pa q_k},
\eeq
we write the equations of motion as
\begin{eqnarray}
X_1:\qquad\qquad q_k^{\bullet}&=&\{q_k,H\}= p_k, \nonumber\\
p_k^{\bullet}&=&\{p_k,H\}=-e^{q_k-q_{k+1}}+e^{q_{k-1}-q_{k}},\qquad\qquad
k=1,\ldots,N-1. \nonumber
\end{eqnarray}
We put $q_{-1}=-\infty,\;  q_N=\infty$ in all formulas. These equations define the vector field $X_1$.

The next, the quadratic  bracket  $\{\; , \; \}_{\pi_1}$ for the Toda was found by M. Adler, \cite{A}. The third, the cubic  bracket
$\{\; , \; \}_{\pi_2}$
was discovered by B. Kupershmidt, \cite{KU}. Contrary to the KdV case and the like the  recurrence operator for the open Toda lattice is not known.

P. Damianou proved, \cite{DA}, that on the  phase space of the open Toda lattice there exists an infinite  sequence of
Poisson brackets
\beq\label{ihpb}
\{\;,\;\}_{\pi_0}\qquad  \{\;,\;\}_{\pi_1}\qquad  \{\;,\;\}_{\pi_2}\qquad .....
\eeq
Any vector field  of  the open Toda  hierarchy can be written using  these brackets and different Hamiltonians  $\HH_0, \HH_1, \HH_2,...$
$$
X_k=\{\;\;,\HH_{k-p}\}_{\pi_p}, \qquad k=1,2,.... ;\qquad  0\leq p \leq k.
$$
 P. Damianou  proved this result  using  implicit  inductive procedure  employing sequence of vector fields which are called master symmetries.
L.  Fayusovich  and M. Gekhhtman, \cite{FG}  associated  this family of Poisson brackets  with multiple Poisson structures on the Riemann sphere.

\vskip 0.5in
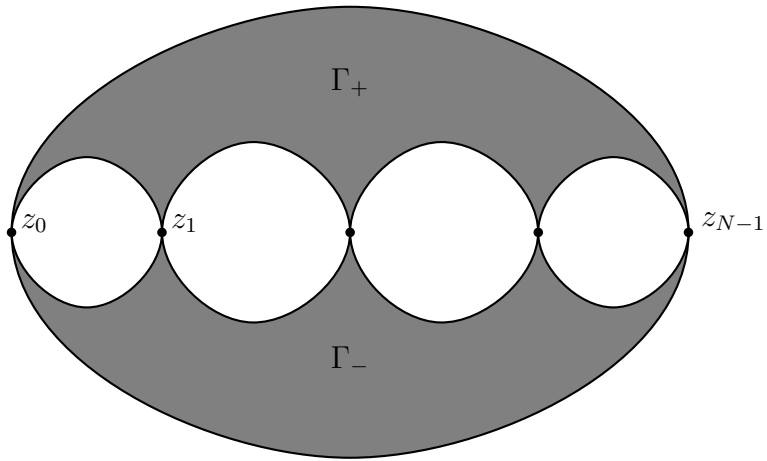
\begin{figure}[tb]
\begin{center}
\begin{pspicture}(0,0)(12,9)
\psccurve[showpoints=false,fillstyle=solid,fillcolor=gray]
(1.5,4.5) (6,7.5)  (10.5, 4.5) (6,1.5)
\psccurve[showpoints=false,fillstyle=solid,fillcolor=white]
(1.5,4.5) (2.5,5.5)  (3.5, 4.5) (2.5,3.5)
\psccurve[showpoints=false,fillstyle=solid,fillcolor=white]
(8.5,4.5) (9.5,5.5)  (10.5, 4.5) (9.5,3.5)
\psccurve[showpoints=false,fillstyle=solid,fillcolor=white]
(3.5,4.5) (4.7,5.7)  (6, 4.5) (4.7,3.3)
\psccurve[showpoints=false,fillstyle=solid,fillcolor=white]
(6,4.5) (7.2,5.7)  (8.5, 4.5) (7.2,3.3)
\psdots(1.5,4.5)(3.5,4.5)(6,4.5)(8.5, 4.5)(10.5, 4.5)
\rput(1.8,4.5){\makebox(0,0)[cb]{$z_0$}}
\rput(3.8,4.5){\makebox(0,0)[cb]{$z_1$}}
\rput(11.1,4.5){\makebox(0,0)[cb]{$z_{N-1}$}}
\rput(6,6.5){\makebox(0,0){$\Gamma_+$}}
\rput(6,2.8){\makebox(0,0){$\Gamma_-$}}


\end{pspicture}
\end{center}
\caption{\small   The reducible Riemann surface $\Gamma$ which consist of two components $\Gamma_-$ and $\Gamma_+$. These components are copies of the Riemann sphere attached to each other at the points $z_0, z_1, \hdots, z_{N-1}$. }
\label{fi:corner}
\end{figure}

\vskip 0.5in

For the open  Toda lattice  the associated spectral curve, the reducible Riemann surface $\Gamma$,  was introduced in \cite{KV}  and presented on Figure 1.
In \cite{V4},  in the context of Nonlinear Schr\"{o}dinger equation,    the author introduced  a new parametrization of the phase space in terms of the associated spectral curve  $\Gamma$  and the Weyl function $\chi=\chi(q), q\in \Gamma;$
$$
\MM \longrightarrow (\Gamma, \chi).
$$
This parametrisation  also  can be constructed  for the open Toda lattice.   For the Toda lattice we used it
to study Poisson bracket $\{\;, \; \}_{\pi_0}$  for the first time   in \cite{V1}.

The goal of the present paper is to write  Poisson brackets for the open Toda in the  explicit  form
$$
\{\chi(p),\chi(q)\}^f= \sum_{k=0}^{N-1} \int\limits_{\overset{\curvearrowright}{O}_k} \omega_{p\, q}^f\; ,
$$
where  the evaluation map is defined as
$$
q:\; (\Gamma, \chi)\rightarrow \chi(q),\qquad\qquad\qquad q\in \Gamma.
$$
The  meromorphic one-form $\omega_{pq}^f$ depends on the functional parameter $f$. When $f$ is  $ n$-th power of the
spectral parameter the formula defines the Poisson bracket $\{\;,\;\}_{\pi_n}$.   The one-form has poles at the points
$z_0, \hdots, z_{N-1};$  at the points $P, Q$  and at infinity. The small circles $O_k$ surround points $z_k$.

An explicit  formula  for   $\omega_{pq}^f$  will be given in the text of the paper. We also demonstrate how to get from this formula
various Darboux systems for the bracket. We construct two such systems. One is associated with poles of the Weyl function and another is associated
with its zeros.

It was conjectured in \cite{V3} that this form is universal.  Namely,  the  formula  of this type
represents  a hierarchy of  Poisson brackets of any integrable hierarchy which can be integrated by  methods of algebraic geometry.
This paper is a confirmation  of this conjecture  for the finite open Toda lattice.

The Sections 2 through 6 are devoted to  the direct spectral transform. We describe  the image  and the range  of the phase space under the direct spectral transform. The image is the space where we construct the Poisson brackets. In Section 7 we present the universal formula for the Poisson brackets on the space of rational functions on the Riemann surface. These PB produce Poisson brackets of the Camassa--Holm equation, \cite{V2},  and magnetic monopoles, \cite{AH}. We present a direct proof of the Jacobi's  identity for these Poisson structures. Due to importance of this result we give two independent  proofs. In Section 8 we construct Dirac's restriction of these Poisson structures  on the image of the direct spectral transform. In Section 9 we present action--angle coordinates. In Section 10 we  describe another system of Darboux's coordinates for this family of brackets.

\section{The commutator formalism.}

Following \cite{F,M},     introduce the new Flaschka--Manakov variables
\bey
c_k &=& e^{q_k-q_{k+1}/2},\quad\qquad\quad\;\;  k=0, 1, \hdots, N-2;\\
v_k &=&  -p_k, \qquad \quad\qquad\qquad k=0, 1, \hdots, N-1.
\eey
The equations of motion take the form
\begin{eqnarray}
X_1:\qquad\qquad  v_k^{\bullet}&=& c_k^2-c_{k-1}^2,\nonumber\\
c_k^{\bullet}&=& c_k(v_{k+1}-v_{k})/2 .\nonumber
\end{eqnarray}
These equations are compatibility conditions  for the Lax equation $L^{\bullet}=\[A,L\]$,  where
$$
L=\left[\begin{array}{ccccc}
v_0      & c_0       & 0         &\cdots        & 0\\
c_0      & v_1       & c_1       &\cdots        & 0\\
\cdot        &  \cdot    & \cdot     &\cdot         & \cdot\\
0    &  \cdots         & c_{N-3}         & v_{N-2}     & c_{N-2}\\
0       & \cdots    & 0         & c_{N-2}      & v_{N-1}
\end{array}\right]
$$
and
$$
2A=\left[\begin{array}{ccccc}
0        & c_0       & 0         &\cdots        & 0\\
-c_0     & 0         & c_1       &\cdots        & 0\\
\cdot        &  \cdot    & \cdot     &\cdot         & \cdot\\
0    &  \cdots         & - c_{N-3}         & 0      & c_{N-2}\\
0       & \cdots    & 0         & - c_{N-2}      &       0
\end{array}\right].
$$
The Lax formula implies that the spectrum $z_0<\ldots < z_{N-1}$ is fixed.

\section{The direct spectral problem.}
 We associate with $L$ the eigenvalue problem
\begin{eqnarray}
v_0y_0 +c_0y_1&=& z y_0, \label{zer}\\
c_{0}y_{0}+v_1y_1 +c_1y_2&=& z y_1, \label{fe}\\
c_{n-1}y_{n-1}+v_ny_n +c_ny_{n+1}&=& z y_n,\qquad n =2,\ldots,N-2;\nonumber \\
c_{N-2}y_{N-2}+v_{N-1}y_{N-1} +c_{N-1}y_{N}&=& z y_{N-1}.\label{la}
\end{eqnarray}
The coefficient $c_{N-1}$ is defined by the formula $c_{N-1} =\prod_{k=0}^{N-2}  c_k^{-1} $.
For the system \ref{zer}--\ref{la} we introduce the solution
$$
P(z):\qquad\qquad P_{-1}(z)=0,\; P_0(z)=1, \ldots,P_N(z);
$$
and for  the system \ref{fe}-\ref{la} another  solution
$$
Q(z):\qquad\qquad  Q_0(z)=0,\; Q_1(z)=\frac{1}{ c_0},\ldots, Q_N(z).
$$

The Weyl function $\chi(z)$ is defined as
$$\chi(z)=-\frac{Q_N(z)}{ P_N(z)};$$
To  state   properties of the function $\chi(z)$ we establish a  determinant   representation of the polynomials $P(z)$ and $Q(z)$.

\begin{lem}
Let $L_{[k,p]}$ be the truncated matrix
\beq\label{mat}
L_{[k,p]}=\left[\begin{array}{ccccc}
v_k      & c_k          &\cdots        & 0\\
c_k      & v_{k+1}       &\cdots        & 0\\
\cdots   & \cdots       & \ddots    &c_{p-1}\\
\cdots   & \cdots       & c_{p-1}      & v_{p}
\end{array}\right].
\eeq
Then for $n=1,2,\ldots,N;$  we have
\beq
P_n(z)=(-1)^n\frac {\det(L_{[0,n-1]}-z I)}{ \prod\limits_{k=0}^{n-1}c_k}
\label{pi},
\eeq
and
\beq
Q_n(z)=(-1)^{n+1}\frac{\det(L_{[1,n-1]}-z I )}{ \prod\limits_{k=0}^{n-1}c_k}.
\label{qu}
\eeq
\end{lem}
\noi
These formulas  can be proved by expanding determinants over the lower row and showing  that  the polynomials satisfy the three term recurrent relation  and the initial conditions.

Formulas \ref{pi}-\ref{qu} imply
\beq\label{we}
\chi(z)=-\frac{(-1)^{N+1}\det (L_{[1,N-1]}-z I)}{ (-1)^N \det (L-z I)}=
\frac{\prod\limits_{s=1}^{N-1} (\gamma_s-z)}{\prod\limits_{n=0}^{N-1} (z_n-z)}=
-\frac{\prod\limits_{s=1}^{N-1} (z-\gamma_s)}{\prod\limits_{n=0}^{N-1} (z-z_n)};
\eeq
where  the roots $z$ and $\g$ interlace
\beq\label{int}
z_{0}<  \g_{1} <  z_{1}  <   \ldots <   z_{N-2} <  \g_{N-1} <  z_{N-1};
\eeq
due to the Sturm theorem, \cite{GK}.
Any  such function  $\chi(z)$  has the form
\beq\label{repr}
\chi(z)=\sum\limits_{k=0}^{N-1} \frac{\rho_k }{  z_k- z};
\eeq
with $\rho_k>0$   and $\sum_{k=0}^{N-1} \rho_k=1$.  For $z$ on the real line and  below/above spectrum the
function $\chi(z)$ is positive/negative.

The functiom $\chi(z)$ can be expanded at infinity as
$$
\chi(z)= -\frac{s_0}{z} - \frac{s_1}{z^2} -\frac{s_2}{z^3}-\hdots,
$$
where $s_p=\sum_{k=0}^{N-1} z_k^p \rho_k,\;  p=0,1,2,...;$  and $s_0=1$.

Following \cite{KK},  consider  functions $\chi(\l)$ with the properties {\it i.}
analytic in the
half--planes $\Im z > 0$ and $\Im z < 0 $. {\it ii.}  $\chi(\bar{z}) = \overline{\chi(z)}$, if
$\Im z \neq 0$
{\it iii.} $\Im \chi(z) > 0, $ if $ \Im z >0$. All such function are called $R$--functions. They play
central role in the spectral theory of selfadjoint operators. The Weyl function of a Jacobi matrix is an $R$--function.

\section{The spaces of rational functions $\Rat_N$  and $\Rat_N'$.}
Consider a connected set  of  functions  $\chi(z)$  on  $\CP^1$   with the property $\chi(\infty)=0$ and $N$  simple  poles at $z_0,z_1,...,z_{N-1}$.
We denote all such functions as $\Rat_N$.
Apparently any function from $\Rat_N$ can be uniquely written as
$$
\chi(z)=-\frac{q(z)}{p(z)}, \quad\quad\text{where}\quad  p(z)= \prod_{k=0}^{N-1} (z-z_k),\quad q(z)=q_0 \prod_{k=1}^{N-1}
(z-\gamma_k).
$$
The space $\Rat_N$  has complex dimension $2N$ and  $z-q(z)$ complex coordinates
$$z_0,...,z_{N-1}; q(z_0),...,q(z_{N-1}).$$
We will need more detailed representation
\beq\label{det}
\chi(z)=-\frac{q_0\prod\limits_{s=1}^{N-1} (z-\gamma_s)}{\prod\limits_{n=0}^{N-1} (z-z_n)}=
-\frac{q_0 z^{N-1} +  q_1 z^{N-2} + \ldots + q_{N-1}}{
z^{N} +  p_0 z^{N-1} + \ldots + p_{N-1}}.
\eeq
Any such function  can be represented as
\beq\label{sf}
\chi(z)=  \sum_{k=1}^{N} \frac{\rho_k}{z_k-z} ,\quad\quad \rho_k= -\R_{z_k} \chi (z).
\eeq
We have another set of $z-\rho$ coordinates
$$z_0,...,z_{N-1}; \rho_0,...,\rho_{N-1}.$$

The function $\chi(z)$ can be expanded at infinity as
$$
\chi(z)= -\frac{s_0}{z} - \frac{s_1}{z^2} -\frac{s_2}{z^3}-\hdots,
$$
where $s_p=\sum_{k=0}^{N-1} z_k^p \rho_k,\;  p=0,1,2,...$.
The polynomials $p(z)$ and $q(z)$ can be reconstructed from this asymptotic expansion if one takes sufficiently many terms of the expansion at infinity.

We denote by $\Rat_N'$ the subset of all functions from $\Rat_N$  which satisfy the condition
$$
q_0=\sum_{n=1}^N \rho_n=1.
$$
The Weyl functions of a Jacobi matrix are exactly those $R$--functions that belong to  $\Rat_N'$.
This implies that  all  $z_k$ are real and $\rho_k >0$  in the representation \ref{sf}.

\section{The Toda lattice hierarchy.}
We denote by $\MM=\RF^N \times \RF^{N-1}_{+}$ the space of all possible $2N-1$ dimensional vectors
$$
(v_0, v_1, \hdots, v_{N-1}; c_0, c_1, \hdots, c_{N-2})
$$
Apparently,  $\MM=\RF^N \times \RF^{N-1}_{+}$.
\subsection{The  hierarchy of Toda flows.}
As was proved by J. Moser, \cite{MO}, that the vector field $X_1$ on the space $\MM$  is the first in the  hierarchy of  commutative  vector fields. In fact,   there   exist  $N$ matricies
$$
A_1=A, A_2, A_3, \hdots , A_N;
$$
which produce a family of commutative vector fields
\beq\label{LE}
X_k: \qquad\qquad \frac{\partial {L}}{\partial t_k}=\[ A_k , {L}\],\qquad\qquad k= 1,2, \hdots, N;
\eeq
and
$$
\frac{\partial \phantom{L}}{\partial t_p}\frac{\partial {L}}{\partial t_k }=\frac{\partial \phantom{L}}{\partial t_k}\frac{\partial {L}}{\partial t_p}.
$$

\begin{thm}\label{dst} \cite{MO}.
The Lax equation  \ref{LE}  implies  that under the action of the vector field $X_k, \; k=1,\hdots, N;$  the parameters $z-\rho$ change according to the rule
\bey
\frac{\partial \phantom{L}}{\partial t_k} \;  z_n&=&0; \\
\frac{\partial \phantom{L}}{\partial t_k}\; \rho_n&=&\(z_n^{k}- \sum_{s=0}^{N-1} z_s^{k} \rho_s\) \rho_n.
\eey
\end{thm}

\subsection{Three  Poisson   structures   on $\MM$.}\label{ttpb}
Any PB bracket on $\MM$   which is    $2N-1$  dimensional  manifold  will be degenerate.	Now we give explicit formulas for the first three PB
together with their Casimirs.

The first  linear bracket $\{\;,\;\}_{\pi_0}$  is the    classical Poisson bracket \ref{cpb} in $c-v$ coordinates
\beq\label{lps}
\{c_k,v_k\}_{\pi_0}=-c_k/ 2,\qquad\qquad\{c_k,v_{k+1}\}_{\pi_0}=c_k/ 2.
\eeq
All other brackets vanish. The functional $\Phi=\tr {L}$ is the Casimir of the bracket.

The second quadratic bracket $\{\;,\;\}_{\pi_1}$  appeared in the work Adler, \cite{A}, and it is defined by the relations
\bey
\{c_k,c_{k+1}\}_{\pi_1}&=&c_k  c_{k+1}/2,\qquad\qquad\{c_k,v_{k}\}_{\pi_1}=-c_k v_k,\\
\{c_k,v_{k+1}\}_{\pi_1}&=&c_k  v_{k+1},\qquad\qquad\{v_k,v_{k+1}\}_{\pi_1}=2 v_k^2.
\eey
All other brackets vanish. The functional $\Phi=\det {L}$ is the Casimir of this  bracket.

The third cubic bracket $\{\;,\;\}_{\pi_2}$  appeared in the work of Kupershmidt, \cite{KU},  and it is defined by the relations
\bey
\{c_k,c_{k+1}\}_{\pi_2}&=&c_k c_{k+1} v_{k+1},\qquad\qquad\quad\;\{c_k,v_{k}\}_{\pi_2}=-c_k v_k^2 - c_k^3,\\
\{c_k,v_{k+1}\}_{\pi_2}&=& c_k v_{k+1}^2 + c_k^3,\qquad\qquad\{c_k,v_{k+2}\}_{\pi_2}=c_k  c_{k+1}^2,\\
\{c_{k+1},v_k\}_{\pi_2}&=&-c_k^2 c_{k+1},\qquad\qquad\quad\;\{v_k,v_{k+1}\}_{\pi_2}=2 c_k^2(v_k+v_{k+1}).
\eey
All other brackets vanish. The functional $\Phi=\tr {L}^{-1}$ is the Casimir of this bracket.

\subsection{The Hamiltonian formulation of the hierarchy of Toda  flows.}
It is known,  \cite{DA},  that the  Poisson brackets introduced above are the  first  three  of the infinite sequence  of compatible Poisson structures
$\pi_k,\; k=0,1,\hdots.$ These higher PB  brackets are constructed in \cite{DA} using  master symmetries.  The construction is based on the inductive procedure.  The general formula for these PB in the $c-v$ coordinates  is not known.

The vector fields 	$X_p$   defined by Lax equations \ref{LE} are  Hamiltonian.  	
Introduce the functions
\beq\label{ham}
\HH_n= \frac{1}{n+1} \tr   { L}^{n+1},\qquad\qquad\qquad n=0,1,\hdots.
\eeq
Then the vector field $X_k$ is produced by the Hamiltonian $\HH_k$ with the Poisson bracket $\{\;,\;\}_{\pi_0}$. Moreover, the vector field  $X_k$ is produced by the Hamiltonian $\HH_{k-p}$ with the Poisson bracket $\{\;,\;\}_{\pi_p}$, where integer $p$ is such that $k \geq p \geq 0.$

\section{The spectral curve and  the new parametrization $(\Gamma, \chi)$.}

Now  we describe  the  direct  spectrum transform from the space of Jacobi matrices $\MM$ to the space of pairs $(\Gamma, \chi )$
$$
\MM  \longrightarrow (\Gamma, \chi ).
$$
The spectral curve $\Gamma=\Gamma_-\bigvee \Gamma_+$ consists of two copies of $\CP^1$ glued together at the points of the spectrum
$z_0, z_1, \hdots, z_{N-1};$ see Figure 1. The points of $\Gamma$ we denote by $q=(z,\pm)$. We define $\chi(q)=\chi(z),\; q\in \Gamma_-$. In fact we deal not
with the whole $\Gamma$ but only with its' $\Gamma_-$ component.

The direct spectral transform is invective. We described its' range  when we described all functions which can appear as Weyl functions of a Jacobi matrix. There are two classical effective ways from the 19th century to invert this map. One  due to  Stieltjes, \cite{GK},  is to expand $\chi$
into a continuous fraction  and another  due to Jacobi, \cite{JA},   is to construct the orthogonal polynomials  $P(z)$ and $Q(z)$ using  the moments $s_p$. The new approach  based on the notion of the BA function for the reducible curve is given in \cite{KV}.

Now we can write a representation of the flows of the Toda hierarchyand PB  in terms of the new parametrization.

\section{The hierarchy of Poisson brackets on $(\Gamma, \chi)$, where $\chi \in \Rat_N$.}
A simplectic  structure on such space of pairs $(\Gamma, \chi)$  was introduced by Atiyah and Hitchin in \cite{AH} as
$$
 \sum_{k=0}^{N-1} \frac{d\, q(z_k)}{q(z_k)} \wedge d\, p(z_k).
$$
The corresponding  Poisson structure is given by the formula
\beq\label{ah}
\{\chi(p), \chi(q)\} =\frac{\(\chi(p) - \chi(q)\)^2}{p-q}.
\eeq
This form was found in the paper of Faybusovich and Gekhtman, \cite{FG}.  For the Atiyah-Hitchin bracket \ref{ah}
it was shown in \cite{V2} that it corresponds  to the main Poisson bracket for  the Camassa-Holm equation written in terms of the Weyl function,
\cite{W},  of the associated Krein's string spectral problem. Faybusovich and Gekhtman also found higher brackets  of the infinite hierarchy of Toda flows  with
\ref{ah} being the first bracket.   In our paper  \cite{GV}  we found an algebraic-geometrical representation of all these  brackets.
It was explained in \cite{GV}  that these brackets produce hierarchy of Poisson brackets of the Camassa-Holm equation.

To introduce a formula for the hierarchy we consider a meromorphic differential $\omega_{p \, q}^{f}$ on   $\Gamma$ which depends on the  entire
function $f(z)$ and two points $p$ and $q$
$$
\omega_{p\, q}^f= \frac{ \epsilon_{p q}(z)}{p -q}\times f(z)  \chi(z) \(\chi(p)-\chi(q)\),
$$
where
$$
\epsilon_{p q}(z)=\frac{1}{2\pi i}\, \[\frac{ 1}{z-p } - \frac{ 1}{z-q}\] dz;
$$
is the standard differential Abelian differential of the third kind with  residues $\pm 1$ at the points $p$ and $q$. It also can be written as
$$
\omega_{p\, q}^f=  \ec_{p q}(z)\times f(z)  \chi(z) \(\chi(p)-\chi(q)\),
$$
where
$$
\ec_{p q}(z)=\frac{1}{2\pi i}\, \[\frac{ 1}{(z-p)(z-q)}\] dz.
$$

The analytic Poisson brackets are defined  on $(\Gamma, \chi),$ where $\chi \in \Rat_N,$ by the formula, see \cite{GV}:
\beq\label{apb}
\{\chi(p),\chi(q)\}^f= \sum_{k=0}^{N-1} \int\limits_{\overset{\curvearrowright}{O}_k} \omega_{p\, q}^f\; ,
\eeq
where the circles $O_k$ are traversed  clockwise and surround points $z_k$.

\begin{thm}\label{JI}
The Poisson bracket  \ref{apb} satisfies the Jacobi identity
$$
\{\{\chi(p), \chi(q)\},\chi(r)\}+ cp\,(p,\,q,\, r)=0.
$$
\end{thm}

We gave an indirect  proof in \cite{GV}  by constructing such  coordinates  that
\ref{apb} has the  standard constant form
\beq\label{conmat}
\newcommand*{\tempi}{\multicolumn{1}{|c}{I}}
\newcommand*{\tempz}{\multicolumn{1}{|c}{0}}
{\mathcal J}=\left[\begin{array}{ccccc}
0      &\tempi \\ \hline
-I      &  \tempz\\
\end{array}
\right].
\eeq
Now we found a direct proof   of Theorem \ref{JI} which is rather difficult.
\subsection{The  first proof of Jacobi identity.}

First we give a proof of the Jacobi identity that  does not use explicit form of  differentials $\epsilon_{pq}(z)$.
Every  we below we omit the superscript $f$ in the formula $\{\;, \;\}= \{\;, \;\}^f$.

 {\it Proof.}  From the definition
$$
\{ \chi(p), \chi(q)\} = \frac{1}{2\pi i} \int\limits_{\bigcup O_k}  \frac{  d z \,f(z) \chi(z) }{(z-p )\, (z-q)}\times   \(\chi(p)-\chi(q)\).
$$
Therefore,
\bey
\{\{ \chi(p), \chi(q)\},\chi(r) \}  &=&  \frac{1}{2\pi i} \int\limits_{\bigcup O_k}  \frac{d z\, f(z) }{(z-p )\, (z-q)}\times \{  \chi(z)\, \(\chi(p)-\chi(q)\), \chi(r) \}\\
&=& \frac{1}{2\pi i} \int\limits_{\bigcup O_k}  \frac{d z\, f(z) \chi(z) }{(z-p )\, (z-q)}\times \{   \(\chi(p)-\chi(q)\), \chi(r) \} \,  +\\
&+& \frac{1}{2\pi i} \int\limits_{\bigcup O_k}  \frac{ d z\, f(z) }{(z-p )\, (z-q)}\times \{  \chi(z), \chi(r) \}\, \(\chi(p)-\chi(q)\)\\
 &=& I + II.  \\
\eey

For the first term we have
\bey
I&=& \frac{1}{2\pi i} \int\limits_{\bigcup O_k}  \frac{  dz f(z)\chi(z)}{(z-p )\, (z-q)}\times \frac{1}{2\pi i}
\int\limits_{\bigcup O_{k'}}  \frac{  d \eta f(\eta) \chi(\eta) }{(\eta-p )\, (\eta-r )} \(\chi(p)-\chi(r)\) \\
&-& \frac{1}{2\pi i} \int\limits_{\bigcup O_k}  \frac{ d z f(z) \chi(z) }{(z-p )\, (z-q)}\times \frac{1}{2\pi i}
\int\limits_{\bigcup O_{k'}}  \frac{  d \eta f(\eta) \chi(\eta) }{(\eta-q )\, (\eta-r )} \(\chi(q)-\chi(r)\)\\
&=& \frac{1}{(2\pi i)^2 } \int\limits_{\bigcup O_k}   \int\limits_{\bigcup O_{k'}}
\frac{ dz d\eta\,f(z)f(\eta)  \chi(z) \chi(\eta)  (\chi(p)-\chi(r)) (z-r) (\eta- q) }  {(  z-p)(z-q)(z-r) (\eta - p)(\eta- r) (\eta-q)}\\
& -& \frac{1}{(2\pi i)^2 } \int\limits_{\bigcup O_k}   \int\limits_{\bigcup O_{k'}}
\frac{ dz d\eta\, f(z)f(\eta) \chi(z) \chi(\eta)  (\chi(q)-\chi(r)) (z-r) (\eta- p)}  {(  z-p)(z-q)(z-r) (\eta - p)(\eta- r) (\eta-q)}.
\eey
Denoting $\PP(z)= (  z-p)(z-q)(z-r), $
\bey
I&=& \frac{1}{(2\pi i)^2 } \int\limits_{\bigcup O_k}   \int\limits_{\bigcup O_{k'}}
\frac{dz d\eta\, f(z)f(\eta)  \chi(z) \chi(\eta)  (\chi(p)-\chi(r)) (z-r) (\eta- q) }  {\PP(z) \PP(\eta) }\\
& -& \frac{1}{(2\pi i)^2 } \int\limits_{\bigcup O_k}   \int\limits_{\bigcup O_{k'}}
\frac{dz d\eta\, f(z)f(\eta)  \chi(z) \chi(\eta)  (\chi(q)-\chi(r)) (z-r) (\eta- p)}  {\PP(z) \PP(\eta) }
\eey
After simple algebra
\bey
I+ c.p. & =& \chi(p) (q-r) \frac{1}{(2\pi i)^2 } \int\limits_{\bigcup O_k}   \int\limits_{\bigcup O_{k'}}
\frac{dz d\eta\,f(z)f(\eta)   \chi(z) \chi(\eta) (\eta +p - 2z)  } {\PP(z) \PP(\eta) } \\
   &+&  \chi(q) (r-p) ... \\
	 &+&  \chi(r) (p-q) ... .
\eey

Similar for the second term we have
\bey
II &=& \frac{1}{2\pi i} \int\limits_{\bigcup O_{k'}}  \frac{ d z f(z)}{(z-p )\, (z-q)} \(\chi(p)-\chi(q)\) \times \frac{1}{2\pi i} \int\limits_{\bigcup O_k} \frac{d \eta f(\eta) \chi(\eta)  }{(\eta-z )\, (\eta-r)} \(\chi(z)-\chi(r)\) \\
&=& \frac{1}{(2\pi i)^2} \int\limits_{\bigcup O_{k'}} \int\limits_{\bigcup O_k}
\frac{d \eta dz\,f(z)f(\eta)  \chi(\eta) \chi(z)\(\chi(p)-\chi(q)\) (\eta - p) (\eta-q)(z-r)}{\PP(z) \PP(\eta) (\eta- z)}   \\
&-& \frac{1}{(2\pi i)^2} \int\limits_{\bigcup O_{k'}} \int\limits_{\bigcup O_k}
\frac{d \eta dz\, f(z)f(\eta) \chi(\eta) \chi(r)\(\chi(p)-\chi(q)\) (\eta - p) (\eta-q)(z-r) }{\PP(z) \PP(\eta) (\eta- z)} \\
&=& A+ B.
\eey
From simple algebra
\bey
B + c.p.&=& \chi(q)\chi(p) (p-q) \frac{1}{(2\pi i)^2} \int\limits_{\bigcup O_{k'}} \int\limits_{\bigcup O_k}
\frac{d \eta dz\,  f(z) f(\eta) \chi(\eta)  (\eta-r)}{\PP(z) \PP(\eta)} \\
&+& \chi(p)\chi(r) (r-p) ...\\
&+& \chi(r)\chi(q) (q-r) ...
\eey
Changing the order of integration
$$
\int\limits_{\bigcup O_{k'}} \int\limits_{\bigcup O_k}
\frac{d \eta dz\,  f(z) f(\eta)\chi(\eta)  (\eta-r)}{\PP(z) \PP(\eta)}=
\int\limits_{\bigcup O_{k'}} \frac{d \eta \,f(\eta) \chi(\eta)  (\eta-r)}{ \PP(\eta)}
\int\limits_{\bigcup O_k}
\frac{ dz f(z) }{\PP(z) }.
$$
The differential $dz f(z)/\PP(z) $  is analytic inside the circles $ O_k$ and the integral vanishes due to the Cauchy theorem.
Therefore,
$$
B + c.p.=0.
$$
This implies
\bey
II + c.p.= A +c.p.&=& \chi(p) (q-r) \frac{1}{(2\pi i)^2 } \int\limits_{\bigcup O_k}   \int\limits_{\bigcup O_{k'}}
\frac{dz d\eta\, f(z)f(\eta)  \chi(z) \chi(\eta) (\eta -p )  } {\PP(z) \PP(\eta) } \\
   &+&  \chi(q) (r-p) ... \\
	 &+&  \chi(r) (p-q) ... .
\eey

Finally,
\bey
I + II + c.p. &=&  \[ \chi(p) (q-r) +  \chi(q) (r-p) + \chi(r) (p-q)\] \times \\
 &\times & \frac{1}{(2\pi i)^2 } \int\limits_{\bigcup O_k}   \int\limits_{\bigcup O_{k'}}
\frac{dz d\eta\,f(z)f(\eta)   \chi(z) \chi(\eta) (2\eta -2z )  } {\PP(z) \PP(\eta) }.
\eey
The last integral vanishes due to skew symmetry.  \qed

\subsection{The  second  proof of Jacobi identity.}

Here we give a second proof of the Jacobi identity that  does not use explicit form of  differentials $\epsilon_{pq}(z)$.

\noindent
 {\it Proof  2.}  From the definition
$$
\{ \chi(p), \chi(q)\} =  \int\limits_{\sum O_k}  \[ \ec_{p q} f\] (z) \times  \chi(z) \(\chi(p)-\chi(q)\).
$$
Therefore,
\bey
\{\{ \chi(p), \chi(q)\},\chi(r) \}  &=&  \int\limits_{\sum O_k} \[ \ec_{p q} f\](z) \times \{  \chi(z)\, \(\chi(p)-\chi(q)\), \chi(r) \}\\
&=& \int\limits_{\sum O_k}  \[\ec_{p q} f\](z)\times \chi(z) \{   \chi(p)-\chi(q), \chi(r) \} \,  +\\
&+&  \int\limits_{\sum O_k} \[\ec_{pq} f\](z) \times \(\chi(p)-\chi(q)\)\, \{  \chi(z), \chi(r) \} \\
 &=& I + II.  \\
\eey

For the first term we have
\bey
I&=& \int\limits_{\sum O_k} \[ \ec_{p q} f \chi\](z)\times \{ \chi(p), \chi(r)\} -
  \int\limits_{\sum O_k}  \[\ec_{p q} f \chi\](z)\times \{ \chi(q), \chi(r)\}\\
&=&  \int\limits_{\sum O_k} \[ \ec_{p q} f \chi\](z)\times
\int\limits_{\sum O_{k'}}   \[ \ec_{p r} f \chi\](\eta) \times   \(\chi(p)-\chi(r)\) \\
& -& \int\limits_{\sum O_k} \[ \ec_{p q} f \chi\](z)\times
\int\limits_{\sum O_{k'}} \[ \ec_{q r} f \chi\](\eta) \times   \(\chi(q)-\chi(r)\)\\
=&+& \chi(p) \int\limits_{\sum O_k} \int\limits_{\sum O_{k'}} \[ \ec_{p q} f \chi\](z)\; \[ \ec_{p r} f \chi\](\eta)\\
&-& \chi(r) \int\limits_{\sum O_k} \int\limits_{\sum O_{k'}} \[ \ec_{p q} f \chi\](z)\; \[ \ec_{p r} f \chi\](\eta)\\
&+& \chi(r) \int\limits_{\sum O_k} \int\limits_{\sum O_{k'}} \[ \ec_{p q} f \chi\](z)\; \[ \ec_{q r} f \chi\](\eta)\\
&-& \chi(q) \int\limits_{\sum O_k} \int\limits_{\sum O_{k'}} \[ \ec_{p q} f \chi\](z)\; \[ \ec_{q r} f \chi\](\eta).
\eey
After simple algebra
\bey
I+ c.p. & =& \chi(p)  \int \int \[ \ec_{p q} f \chi\](z)\; \[ \ec_{p r} f \chi\](\eta) -
           \[ \ec_{rp} f \chi\](z)\; \[ \ec_{p q} f \chi\](\eta) \\
    &&\qquad\quad\;\;  -\[ \ec_{qr} f \chi\](z)\; \[ \ec_{qp} f \chi\](\eta) +\[ \ec_{qr} f \chi\](z)\; \[ \ec_{rp} f \chi\](\eta) \\
   &+&  \chi(q)  ... \\
	 &+&  \chi(r)  ... .
\eey
Using the first identity
\beq\label{firid}
\frac{\ec_{ab}(z) }{z-c}= \frac{\ec_{a' b'}(z) }{z-c'},
\eeq
where $(a',b',c')$ is an arbitrary permutation of the points $(a,b,c)$,
and the second identity
$$
(z-r)(\eta-q)-(z-q)(\eta-r)-(z-p)(\eta-r)+(z-p)(\eta-q)=(\eta+p -2z)(q-r),
$$
we  transform the expression under integral sign to the form
\bey
I+ c.p. & =& \chi(p) (q-r) \int \int  \frac{\[\ec_{p q}f \chi\](z)  \[\ec_{p r}f \chi\](\eta)}{(z-r)(\eta-q)} (\eta+p -2z) \\
              &+&  \chi(q)  ... \\
	 &+&  \chi(r)  ... .
\eey

Similar for the second term we have
\bey
II &=& \(\chi(p)-\chi(q)\)\times \,  \int\limits_{\sum O_k} \[ \ec_{pq} f\](z)
\int\limits_{\sum O_{k'}}  \[\ec_{z r} f \chi\](\eta) \times   \(\chi(z)-\chi(r)\) \\
&=& \(\chi(p)-\chi(q)\)\times \,  \int\limits_{\sum O_k}  \[\ec_{pq} f \chi\] (z)
\int\limits_{\sum O_{k'}}  \[ \ec_{z r} f \chi\](\eta)    \\
&-& \(\chi(p)-\chi(q)\) \chi(r)\times \,  \int\limits_{\sum O_k}  \[ \ec_{pq} f\](z)
\int\limits_{\sum O_{k'}}  \[ \ec_{z r} f \chi\](\eta)    \\
&=& A-  B.
\eey

It is easy to see
\bey
A + c.p.&=& \chi(p) \[ \quad  \int\limits_{\sum O_k}   \[\ec_{pq} f \chi\](z)
\int\limits_{\sum O_{k'}}  \[ \ec_{z r} f \chi\](\eta)
 - \int\limits_{\sum O_k}   \[\ec_{rp} f\](z)
\int\limits_{\sum O_{k'}}  \[ \ec_{zq} f \chi \] (\eta)   \]\\
&+& \chi(q)  ...\\
&+& \chi(r)  ...  .
\eey
Using the identity
$$
(z-r)(\eta-p)(\eta-q)- (z-q)(\eta-p)(\eta-r)=(\eta - p)(r-q)(z-\eta)
$$
we obtain
\bey
A + c.p.&=& \chi(p) (r-q)    \int\limits_{\sum O_k} \int\limits_{\sum O_{k'}}
\frac{\[\ec_{pq} f \chi\](z)   \[ \ec_{z r} f \chi\](\eta)} {(z-r)(\eta-q)} \, (z-\eta) \\
&+& \chi(q)  ...\\
&+& \chi(r)  ...  .
\eey
Using the identity
$$
\frac{   \ec_{z r} (\eta)} {\eta-q} \, (z-\eta) =-\ec_{r q},
$$
we have
\bey
A + c.p.&=& \chi(p) (q-r)    \int\limits_{\sum O_k} \int\limits_{\sum O_{k'}}
\frac{\[\ec_{pq} f \chi\](z)   \[ \ec_{ r q} f \chi\](\eta)} {(z-r)(\eta-p)} \, (\eta-p) \\
&+& \chi(q)  ...\\
&+& \chi(r)  ...  .
\eey

From simple algebra
\bey
B + c.p.&=& \chi(q)w(p) \[ \quad  \int\limits_{\sum O_k}   \[\ec_{qr} f\](z)
\int\limits_{\sum O_{k'}}  \[ \ec_{z p} f \chi\](\eta)
 - \int\limits_{\sum O_k}   \[\ec_{r p} f\](z)
\int\limits_{\sum O_{k'}}  \[ \ec_{z q} f \chi \] (\eta)   \]\\
&+& \chi(p)\chi(r)  ...\\
&+& \chi(r)\chi(q)  ...
\eey
We are going to transform the expression in the square bracket
using the first identity  \ref{firid} and the second identity
$$
(z-p)(\eta-q)(\eta-r)- (z-q)(\eta-p)(\eta-r)=(\eta-r)(z-\eta)(p-q).
$$
Therefore,
\bey
B + c.p.&=& \chi(q)\chi(p)
\times  \[ \quad \int\limits_{\sum O_k} \int\limits_{\sum O_{k'}}
\frac{\[ \ec_{qr} f\](z)  \[\ec_{z p} f \chi\](\eta) (\eta-r)(z-\eta)(p-q)}{(z-p)(\eta-q)(\eta-r)}
\] \\
&+& \chi(p)\chi(r)  ...\\
&+& \chi(r)\chi(q)  ...
\eey
Note,
$$
\ec_{z p} (\eta) (z-\eta)=-\frac{d \eta}{\eta -p}=\ec_{p\infty} (\eta).
$$
Changing the order of integration
\bey
\quad \int\limits_{\sum O_k} \int\limits_{\sum O_{k'}}&&
\frac{ \[\ec_{qr} f\](z)  \[\ec_{z p} f \chi\](\eta) (\eta-r)(z-\eta)(p-q)}{(z-p)(\eta-q)(\eta-r)}
\\
&=&\int\limits_{\sum O_{k'}} \frac{ \[\ec_{p \infty}   f \chi\](\eta) (\eta-r)(p-q)}{(\eta-q)(\eta-r)}
\int\limits_{\sum O_k} \frac{ \[\ec_{qr} f\](z) }{z-p}
\eey
The differential  is analytic inside the circles $ O_k$ and the integral vanishes due to the Cauchy theorem.
Therefore,
$$
B + c.p.=0.
$$
This implies
\bey
II + c.p.= A +c.p.&=& \chi(p) (q-r)    \int\limits_{\sum O_k} \int\limits_{\sum O_{k'}}
\frac{\[\ec_{pq} f \chi\](z)   \[ \ec_{ r q} f \chi\](\eta)} {(z-r)(\eta-p)} \, (\eta-p)  \\
   &+&  \chi(q) (r-p) ... \\
	 &+&  \chi(r) (p-q) ... .
\eey

Finally,
\bey
I + II + c.p. =  \[ \chi(p) (q-r) +  \chi(q) (r-p) + \chi(r) (p-q)\] \times \\
\times\int\limits_{\sum O_k} \int\limits_{\sum O_{k'}}  \frac{\[\ec_{pq} f \chi\](z)   \[ \ec_{ r q} f \chi\](\eta)} {(z-r)(\eta-p)} \, (2\eta-2z).
\eey
The last integral vanishes due to skew symmetry.  \qed

\subsection{Two quadratic algebras.}
By the Cauchy formula from \ref{apb} we have for any entire $f(z)$
\bey
\{\chi(p),\chi(q)\}^f & = & \R_p \, \omega_{p\, q}^f + \R_q \, \omega_{p\, q}^f +\R_{\infty} \, \omega_{p\, q}^f \\
                & = & \frac{f(p) \chi(p)-f(q)\chi(q)}{p-q}\(\chi(p)-\chi(q)\)+
\R_{\infty} \, \omega_{p\, q}^f.
\eey
If $f(z)=z^n,\, n=0,1,\hdots;$ then the residue at infinity vanishes identically only for $n=0$ or $1.$
When $f(z)=1$ we obtain quadratic Poisson algebra corresponding to the {\it rational} solution of CYBE
\beq\label{rpb}
\{\chi(p),\chi(q)\}^1=\(\chi(p)-\chi(q)\)\frac{\chi(p)-\chi(q)}{p-q}.
\eeq
Another quadratic Poisson algebra is obtained for $f(z)=z$ and it corresponds to the {\it trigonometric} solution of CYBE
\beq\label{tpb}
\{\chi(p),\chi(q)\}^z=\(p\chi(p)-q\chi(q)\)\frac{ \chi(p)- \chi(q)}{p-q}.
\eeq
It can be verified directly that \ref{rpb} and \ref{tpb}  satisfy Jacobi identity.

 \subsection{The Darboux coordinates.}
The following  result  is proved  using residues:
\begin{thm}
 (\cite{GV}, Theorem 2.1) The Poisson bracket  \ref{apb}  in $z-\rho$ coordinates has the form
\begin{eqnarray}
\{\rho_k,\rho_n\}&=&\frac{(f(z_k) +f(z_n)) \rho_k \,\rho_n}{ z_n-z_k}(1-\delta_k^n),\label{rr}\\
\{\rho_k,z_n\}&=& \rho_k  f(z_n) \delta_k^n, \label{rl}\\
\{z_k,z_n\}&=& 0 \label{ll}.
\end{eqnarray}
\end{thm}

This result together with the formula $q(z_k)= p'(z_k) \rho_k$ implies that \ref{apb}  in $z-q(z)$ coordinates takes the form
$$
\{q(z_k), z_n\}= f(z_n) q(z_k) \delta_k^n;
$$
and all other brackets vanish
$$
\{q(z_k), q(z_n)\}=\{z_k, z_n\}=0.
$$
In  the cooordinates
\beq\label{xk}
I_k=\int^{z_k}_{\infty} \frac{d \zeta} {f(\zeta)},\qquad k=0,1,\hdots, N-1;
\eeq
and
$$
y_k=\ln q(z_k),\qquad k=0,1,\hdots, N-1.
$$
We define  $\infty$  some  fixed point on the  Riemann  sphere.  We  need a convergence of the integral.
The bracket \ref{apb} takes the standard constant form
$$
\{y_k, I_n\}= \delta_k^n;\qquad\qquad
\{y_k, y_n\}=\{I_k, I_n\}=0.
$$

\section{The hierarchy of Poisson brackets on  $(\Gamma, \chi)$, where $\chi \in \Rat_N'$.}

\begin{lem}\label{pbqr}  For  two functionals
$$
\Phi_1=I_0+I_1+\hdots + \hdots + I_{N-1},\qquad\qquad\qquad \Phi_2= \log q_0;
$$
where $I_k$    are defined by \ref{xk},   the bracket \ref{apb}  is
$$
\{\Phi_1, \Phi_2\}^f=1.
$$
\end{lem}

\noindent
 {\it Proof.}  Defined function $F(z)$ by the formula
$$
F(z)= \int\limits_{\infty}^{z} \frac{ds}{f(s)}.
$$
We assume that zeros of $f(z)$ are distinct from the poles $z_k$.  Therefore, $F(z)$ is well defined in the vicinity of the polls.
Integrating by parts
\bey
\Phi_1&=&I_0+I_1+\hdots + \hdots + I_{N-1}=F(z_1) + F(z_2) +\hdots +  F(z_{N-1})\\
&=& \sum_{k} -\frac{1}{2\pi i} \int\limits_{O_k} F(\zeta) d \ln \chi(\zeta)\\
 &=& \sum_{k} \frac{1}{2\pi i} \int\limits_{O_k} \frac{1}{f(\zeta)}  \ln \chi(\zeta)  d\zeta.
\eey
Using  $q_0=- \lim_{y\rightarrow \infty} y \chi(y)$,  we have
\bey
\{\Phi_1, q_0\}&=& - \lim_{y\rightarrow \infty} y \{\Phi_1, \chi(y)\}=-\sum_{k} \frac{1}{2\pi i} \int\limits_{O_k} \frac{1}{f(\zeta) \chi(\zeta)}
\lim_{y\rightarrow \infty} y \{\chi(\zeta), \chi(y)\}  d\zeta.
\eey
Using \ref{apb},  the Cauchy's formula for sufficiently large $R$
\bey
\lim_{y\rightarrow \infty} y \{\chi(\zeta), \chi(y)\} &=&\lim_{y\rightarrow \infty}  \sum_{k'}  \frac{1}{2\pi i} \int\limits_{O_{k'}}
\frac{y \, dz}{(z-\zeta)(z-y)}  f(z) \chi(z)  (\chi(\zeta)- \chi(y))\\
 &=&-\chi(\zeta)  \sum_{k'}  \frac{1}{2\pi i} \int\limits_{O_{k'}}
\frac{\, dz}{z-\zeta}  f(z) \chi(z)\\
 &=&-   f(\zeta) \chi^2(\zeta) + \chi(\zeta) \frac{1}{2\pi i} \int\limits_{O_R}
\frac{\, dz}{z-\zeta}  f(z) \chi(z) = I +II.
\eey
Changing the order of integration  for any $k$ we see that the integral vanishes due to the Cauchy theorem
$$
-  \int\limits_{O_k} \frac{d\zeta}{f(\zeta)}  \int\limits_{O_R}
\frac{\, dz}{z-\zeta}  f(z) \chi(z) = \int\limits_{O_R}  dz  f(z) \chi(z)  \int\limits_{O_k} \frac{d\zeta}{( z-\zeta  ) f(\zeta)} =0.
$$
Therefore, contribution of the  term  $II$  is zero.  For the  term $I$   we have
\bey
\{\Phi_1, q_0\}&=& - \lim_{y\rightarrow \infty} y \{\Phi_1, \chi(y)\}=\sum_{k} \frac{1}{2\pi i} \int\limits_{O_k} \frac{d \zeta }{f(\zeta) \chi(\zeta)}  f(\zeta) \chi^2(\zeta)\\
&=&\sum_{k} \frac{1}{2\pi i} \int\limits_{O_k}  \chi(\zeta)d \zeta = \frac{1}{2\pi i} \int\limits_{O_R}  \chi(\zeta) d \zeta= q_0, \\
\eey
where $R$ is sufficiently large number.
\qed

\begin{thm}
For any choice of an entire function $f(z)$  a Dirac restriction of  the Poisson bracket \ref{apb} on the sub-manifold
\beq\label{subman}
\Phi_1=c_1,\qquad\qquad\qquad \Phi_2=c_2;
\eeq
where $x_k$    are defined by \ref{xk},  is given by the  formula
\beq\label{apbr}
\{\chi(p),\chi(q)\}^f= \sum_{k=0}^{N-1} \int\limits_{\overset{\curvearrowright}{O}_k} \tilde{\omega}_{p\, q}^f\; ,
\eeq
where the circles $O_k$
 are traversed  clockwise and surround points $z_k.$  The  new modified differential $\tilde{\omega}_{p\, q}^f$  is
$$
\tilde{\omega}_{p\, q}^f= \frac{ \epsilon_{p q}(z)}{p -q}\times f(z)  \chi(z) \(\chi(p)-\chi(q)\)-
 \epsilon_{p q}(z)\times f(z)  \chi(z) \chi(p)\chi(q)  e^{-c_2}.
$$
\end{thm}

\noindent
 {\it Proof.}   According to the Dirac's recipe, \cite{DI},  we modify the original bracket  $\{\;,\;\}$  by adding  two extra terms
$$
F_1^{\bullet} = \{F_1,F_2\}'= \{F_1,F_2\} +\sigma_1 \{F_1, \Phi_1\} +\sigma_2\{F_1, \Phi_2\}.
$$
The constants $\sigma_1$ and $\sigma_2$  are specified by conditions
\bey
\Phi_1^{\bullet} &=& \{\Phi_1,F_2\}'=   \{\Phi_1,F_2\} +\sigma_1 \{\Phi_1, \Phi_1\} +\sigma_2\{\Phi_1, \Phi_2\}=0,\\
 \Phi_2^{\bullet} &=& \{\Phi_2,F_2\}'=   \{\Phi_2,F_2\} +\sigma_1 \{\Phi_2, \Phi_1\} +\sigma_2\{\Phi_2, \Phi_2\}=0.
\eey
Using Lemma \ref{pbqr}   we have
$$
\sigma_1=\{\Phi_2,F_2\} \,\qquad\qquad\qquad \sigma_2=-\{\Phi_1,F_2\},
$$
and
$$
 \{F_1,F_2\}'= \{F_1,F_2\} +\{\Phi_2,F_2\}  \{F_1, \Phi_1\} -\{\Phi_1,F_2\}  \{F_1, \Phi_2\}.
$$
Therefore,
$$
 \{\chi(x),\chi(y)\}'= \{\chi(x),\chi(y)\} +\{\Phi_2,\chi(y)\}  \{\chi(x), \Phi_1\} -\{\Phi_1,\chi(y)\}  \{\chi(x), \Phi_2\}.
$$
Using,
\beq\label{hhh}
\{\Phi_1, \chi(y)\}= \chi(y);
\eeq
and
\beq\label{jjj}
\{\Phi_2,\chi(y)\}=-\chi(y) e^{-\Phi_2}  \sum_{k} \frac{1}{2\pi i} \int\limits_{O_k} \frac{d \zeta}{\zeta-y} f(\zeta) \chi(\zeta).
\eeq
we obtain the result.

To prove \ref{hhh} we use formula for $\Phi_1$ obtained in the proof of Lemma \ref{pbqr}
\bey
\Phi_1  &=& \sum_{k} \frac{1}{2\pi i} \int\limits_{O_k} \frac{1}{f(\zeta)}  \ln \chi(\zeta)  d\zeta.
\eey
Therefore,
\bey
\{\Phi_1, \chi(y)\}&=&\sum_{k} \frac{1}{2\pi i} \int\limits_{O_k} \frac{1}{f(\zeta)\chi(\zeta)}  \{\chi(\zeta), \chi(y)\}   d\zeta \\
&=& \sum_{k} \frac{1}{2\pi i} \int\limits_{O_k} \frac{ d\zeta }{f(\zeta)\chi(\zeta)}  \sum_{k'} \frac{1}{2\pi i}
\int\limits_{O_{k'}} \frac{d \eta}{(\eta-\zeta)(\eta- y)}  f(\eta) \chi(\eta) (\chi(\zeta)- \chi(y))
\eey
When $k \neq k'$ we have
$$
\int\limits_{O_k} \frac{d \zeta (\chi(\zeta)-\chi(y)) }{f(\zeta) \chi(\zeta)} \int\limits_{O_{k'}} \frac{d \eta} {(\eta- \zeta)(\eta-y)} f(\eta) \chi(\eta)=0
$$
due to the Cauchy theorem.  When $k=k'$ we have changing the order of the  integration
\bey
\therefore&=&\sum_k \frac{1}{2\pi i}\int\limits_{O_k} \frac{d \zeta (\chi(\zeta)-\chi(y)) }{f(\zeta) \chi(\zeta)} \frac{1}{2\pi i}\int\limits_{O_{k}} \frac{d \eta} {(\eta- \zeta)(\eta-y)} f(\eta) \chi(\eta)\\
&=&\sum_k \frac{1}{2\pi i}\int\limits_{O_{k}} \frac{d \eta} {(\eta-y)} f(\eta) \chi(\eta)\frac{1}{2\pi i}\int\limits_{O_k} \frac{d \zeta (\chi(\zeta)-\chi(y)) }{f(\zeta) \chi(\zeta)(\eta- \zeta)}\\
&=& \sum_k \frac{1}{2\pi i}\int\limits_{O_{k}} \frac{\chi(\eta)d\eta} {\eta-y} =\frac{1}{2\pi i}\int\limits_{O_{R}} \frac{\chi(\eta)d\eta} {\eta-y}=\chi(y).
\eey

To prove \ref{jjj}, we have
\bey
\{q_0, \chi(y)\}&=& - \{   \lim_{x\rightarrow \infty}  x \chi(x), \chi(y)\}\\
&=&- \lim_{x\rightarrow \infty}  \sum_{k} \frac{1}{2 \pi i} \int\limits_{O_k} \frac{x d\zeta}{(\zeta-x)(\zeta-y)} f(\zeta) \chi(\zeta) (\chi(x) -\chi(y))\\
&=& -\chi(y)  \sum_{k} \frac{1}{2 \pi i} \int\limits_{O_k} \frac{ d\zeta}{\zeta-y} f(\zeta) \chi(\zeta).
\eey
This implies the result. \qed

As a byproduct  we have that the  bracket  defined by \ref{apbr} satisfies the Jacobi identity.
This statement  also can be proved directly,  similar to Theorem \ref{JI}.

The bracket   \ref{apbr} has two Casimirs $\Phi_1$ and  $\Phi_2$  and,  therefore, degenerates  on $\Rat_N$ with  the  rank $2N-2$.
The bracket   \ref{apbr}   produces  vector fields  in $\Rat_N$  that are  tangent to $\Rat_N'$  and  therefore  it can be restricted to this subspace.   We call it a bracket on   $(\Gamma, \chi)$, where $\chi \in \Rat_N'$.

For $f(z)=1$  we have
$$
\Phi_1=z_0+z_1+\hdots + \hdots + z_{N-1}=p_0.
$$
 For $f(z)=z$  we have
$$
\Phi_1=\ln z_0+\ln z_1+\hdots + \ln z_{N-1}=\ln p_{N-1}.
$$
For $f(z)=z^{n+1},\; n=1,2,\hdots; $  we have
$$
\Phi_1=\frac{1} {z_0^n}+\frac{1} {z_1^n} +\hdots  +\frac{1} {z_{N-1}^n}.
$$
These are the Casimirs of    three Poisson structures considered in Section \ref{ttpb}.

\subsection{The Toda flows in terms of $(\Gamma,\chi)$.}
Now we compute brackets \ref{apbr} in terms of the coordinates $z-\rho$ similar to    Theorem 2.1 in \cite{GV}.
\begin{thm}\label{rrr}
 The Poisson bracket  \ref{apbr}  in $z-\rho$ coordinates has the form
\begin{eqnarray}
\{\rho_k,\rho_n\}^f&=& \( f(z_k) + f(z_n)\)  \times  (1-\delta_n^k)\times   \\ \nonumber
                                 &\times &  \[ \frac{\rho_k \rho_n}{z_k -z_n}   - \rho_k \rho_n
\( \sum_{m\neq k} \frac{\rho_m }{z_m -z_k}  -\sum_{m\neq n} \frac{\rho_m }{z_m -z_n} \)  \],\label{rr}\\
\{\rho_k,z_n\}^f&=&   \rho_k f(z_n) \delta^n_k-   f(z_n) \rho_k \rho_n            , \label{rl}\\
\{z_k,z_n\}^f&=& 0.   \label{ll}
\end{eqnarray}
\end{thm}

Now consider Hamiltonians defined by \ref{ham}
$$
\HH_n=\frac{1}{n+1} \tr { L}^{n+1}=\frac{1}{n+1} \sum_{k=0}^{N-1} z_k^{n+1}, \qquad\qquad   n=0,1,\hdots.
$$
Due to Theorem \ref{rrr}, one has for the brackets $\{\;,\;\}^{z^p}$ and integer $k, p:\; k\geq p \geq 0;$
$$
z_n^{\bullet}= \{z_n, \HH_{k-p}\}^{z^p}=0
$$
and
$$
\rho_n^{\bullet}= \{\rho_n, \HH_{k-p}\}^{z^p}=\(z_n^{k}- \sum_{s=0}^{N-1} z_s^{k} \rho_s\) \rho_n.
$$
These  equations coincide with the equations for the $X_{k}$ vector field   in  Theorem  \ref{dst}.

\subsection{Two quadratic algebras.}
As before,   from \ref{apbr} for any entire $f(z)$  we have
$$
\{\chi(p),\chi(q)\}^f  =  \R_p \,  \tilde{ \omega}_{p\, q}^f + \R_q \,  \tilde{ \omega}_{p\, q}^f +\R_{\infty} \,  \tilde{ \omega}_{p\, q}^f
$$
$$
=\(f(p) \chi(p)-f(q)\chi(q)\) \(\frac{\chi(p)-\chi(q)}{p-q}  -\chi(p) \chi(q)e^{-c_2}\)
+ \R_{\infty} \, \tilde{ \omega}_{p\, q}^f.
$$
When $f(z)=1$ we obtain quadratic Poisson algebra
\beq\label{rpbr}
\{\chi(p),\chi(q)\}^1=\(\chi(p)-\chi(q)\)\(\frac{\chi(p)-\chi(q)}{p-q}- \chi(p) \chi(q)e^{-c_2}\).
\eeq

For $f(z)=z$  we have
\beq\label{tpbr}
\{\chi(p),\chi(q)\}^z=\(p\chi(p)-q\chi(q)\)\(\frac{\chi(p)-\chi(q)}{p-q}- \chi(p) \chi(q)e^{-c_2}\).
\eeq
When $f(z)=z^n,\, n=0,1,\hdots;$ then the residue at infinity vanishes identically only for $n=0$ or $1.$ This implies that higher Toda brackets with
$n \geq 2$ do not form the quadratic Poisson algebra and therefore do not admit $R$--matrix representation.

\section{The action-angle coordinates}

In  the cooordinates
$$
I_k=\int^{z_k}_{\infty} \frac{d \zeta} {f(\zeta)};\qquad\qquad \theta_k=\ln \frac{(-1)^k q(z_k)}{q(z_0)},\qquad k=1,\hdots, N-1.
$$
where  $\infty$ is a fixed point on the  Riemann  sphere,   the bracket \ref{apbr} takes the standard constant form.  As it is explained in \cite{V1}  the variables  $\theta_k$ are angles  on the isospectral manifold parametrized  by the non-standard  Abel  map constructed from the differentials of the third type on  the spectral curve $\Gamma$.
The ispectral manifold is not compact and the angles are not a cyclic variables; the vector $(\theta_1,\hdots,\theta_{N-1})$ takes value in $R^{N-1}$.

The following theorem was proved in  \cite{V1} for $f=1$ .  The methods of this paper extend easiliy to the case of general  $f$.

\begin{thm}
The Poisson bracket  \ref{apbr}  in $I-\theta$ coordinates has the form
\begin{eqnarray}
\{\theta_k, \theta_n\}^f&=&0,\label{rr}\\
\{I_k, I_n\}^f&=& 0, \label{rl}\\
\{\theta_k, I_n\}^f&=& \delta_k^n  \label{ll}.
\end{eqnarray}
\end{thm}

\section{Another system of Darboux's coordinates} In this section we construct  system of Darboux coordinates  for \ref{apbr}  on $\Rat_N$ associated
with zeroes of   these  functions. Let us define
$$
\pi_k=\log (-1)^{N+k}  p(\gamma_k), \qquad\qquad\qquad k=1,\hdots, N-1.
$$
In additions to $\Phi_1$ and $\Phi_2$ we use
$$
\gamma_1,\hdots, \gamma_{N-1}; \qquad\qquad\qquad \pi_1,\hdots , \pi_{N-1}.
$$
The following result   was proved (Theorem 3  in \cite{V1})  for $f=1$.  Now we can formulate it for general $f$.
\begin{thm}
 The Poisson bracket  \ref{apb}  in $\gamma-\pi$ coordinates has the form
\begin{eqnarray}
\{\gamma_k, \pi_n \}^f&=&\delta_k^n,\label{rr}\\
\{\Phi_1, \Phi_2 \}^f&=& 1\label{rl}.
\end{eqnarray}
All other brackets vanish.
\end{thm}

From this the  relations \ref{rr} follow easily for the restricted bracket \ref{apbr}.

\vskip .1in
\noindent
\newline
Department of Mathematics
\newline
Michigan State University
\newline
East Lansing, MI 48824
\newline
USA
\vskip 0.1in
\noindent
vaninsky@math.msu.edu

\end{document}